\documentstyle[twoside,bezier,12pt]{article}

\catcode`\@=11
\long\def\@makefntext#1{ 
\protect\noindent \hbox to 3.2pt {\hskip-.9pt
$^{{\eightrm\@thefnmark}}$\hfil}#1\hfill} 

\def\thefootnote{\fnsymbol{footnote}}
 \def\@makefnmark{\hbox to 0pt{$^{\@thefnmark}$\hss}}  

\def\ps@myheadings{\let\@mkboth\@gobbletwo
\def\@oddhead{\hbox{} 
\rightmark\hfil\eightrm\thepage}
\def\@oddfoot{}\def\@evenhead{\eightrm\thepage\hfil 
\leftmark\hbox{}}\def\@evenfoot{}
\def\sectionmark##1{}\def\subsectionmark##1{}}

\oddsidemargin=\evensidemargin
\renewcommand{\thefootnote}{\fnsymbol{footnote}}

\newcounter{sectionc}\newcounter{subsectionc}\newcounter{subsubsectionc}
\renewcommand{\section}[1] {\vspace{12pt}\addtocounter{sectionc}{1}
\setcounter{subsectionc}{0}\setcounter{subsubsectionc}{0}\noindent
	{\bf\thesectionc. #1}\par\vspace{5pt}}
\renewcommand{\subsection}[1] {\vspace{12pt}\addtocounter{subsectionc}{1}
	\setcounter{subsubsectionc}{0}\noindent
	{\bf\thesectionc.\thesubsectionc. {\kern1pt \bf\it #1}}\par\vspace{5pt}}
\renewcommand{\subsubsection}[1] {\vspace{12pt}\addtocounter{subsubsectionc}{1}
	\noindent{\thesectionc.\thesubsectionc.\thesubsubsectionc.
	{\kern1pt \it #1}}\par\vspace{5pt}}

\newcommand{\textlineskip}{\baselineskip=14pt}

\def\eightcirc{
\begin{picture}(0,0)
\put(4.4,1.8){\circle{6.5}}
\end{picture}}
\def\eightcopyright{\eightcirc\kern2.7pt\hbox{\eightrm c}}

\def\abstracts#1#2#3{{
	\centering{\begin{minipage}{5in}\baselineskip=12pt\tenrm
	\centerline{ABSTRACT}
	\parindent=0pt #1\par
	\parindent=15pt #2\par
	\parindent=15pt #3
	\end{minipage} }\par}}

\newcommand{\bibit}{\it}
\newcommand{\bibbf}{\bf}
\renewenvironment{thebibliography}[1]			
	{
	 \begin{list}{\arabic{enumi}.}			
	{\usecounter{enumi}\setlength{\parsep}{0pt}
	 \setlength{\leftmargin 17pt}{\rightmargin 0pt}	
	 \setlength{\itemsep}{0pt} \settowidth		
	{\labelwidth}{#1.}\sloppy}}{\end{list}}	

\newcounter{itemlistc}
\newcounter{romanlistc}
\newcounter{alphlistc}
\newcounter{arabiclistc}

\newcounter{tempfigtabc}			

\newcounter{temptabtabc}
\newcommand{\tcaption}[1]{			
	\setcounter{temptabtabc}{\thetable}
	\addtocounter{temptabtabc}{1}
     {\noindent\parbox{6truein}{\tenrm Table~\thetemptabtabc. #1} }}

\def\pmb#1{\setbox0=\hbox{#1}
	\kern-.025em\copy0\kern-\wd0
	\kern.05em\copy0\kern-\wd0
	\kern-.025em\raise.0433em\box0}

\def\fnt#1#2{\footnotetext{\kern-.3em
	{$^{\mbox{\scriptsize #1}}$}{#2}}}

\def\fpage#1{\begingroup
\voffset=.3in
\thispagestyle{empty}\begin{table}[b]\centerline{\footnotesize #1}
	\end{table}\endgroup}

\def\runninghead#1#2{\pagestyle{myheadings}
\markboth{{\eightit{\quad #1}}\hfill}{\hfill{\eightit{#2\quad}}}}

 1
 1
 1

\font\eightrm=cmr8
\font\eightit=cmti8

\newcommand{\br}{\mbox{\boldmath $r$}}

\newcommand{\bpi}{\mbox{\boldmath $\pi$}}

\newcommand{\btau}{\mbox{\boldmath $\tau$}}

\newcommand{\bepsilon}{\mbox{\boldmath $\epsilon$}}

\newcommand{\bsigma}{\mbox{\boldmath $\sigma$}}
\newcommand{\sbtau}{\mbox{\footnotesize {\boldmath $\tau$}}}

\newcommand{\ba}{\begin{eqnarray}}
\newcommand{\ea}{\end{eqnarray}}
\newcommand{\be}{\begin{equation}}
\newcommand{\ee}{\end{equation}}
\newcommand{\trom}{{\rm \, tr \,}}

\def\qed{\hbox{${\vcenter{\vbox{                          
   \hrule height 0.4pt\hbox{\vrule width 0.4pt height 6pt
   \kern5pt\vrule width 0.4pt}\hrule height 0.4pt}}}$}}

\runninghead{}{}

\begin{document}
\normalsize\textlineskip
{\thispagestyle{empty}
\setcounter{page}{1}

\renewcommand{\thefootnote}{\fnsymbol{footnote}} 
\def\bsc{{\sc a\kern-6.4pt\sc a\kern-6.4pt\sc a}}
\def\bflatex{\bf L\kern-.30em\raise.3ex\hbox{\bsc}\kern-.14em
T\kern-.1667em\lower.7ex\hbox{E}\kern-.125em X}

\fpage{1}
\centerline{\bf THE LOW-ENERGY THEOREM OF PION PHOTOPRODUCTION}
\centerline{\bf IN SOLITON MODELS OF THE NUCLEON}
\vspace*{0.035truein}

\vspace{0.37truein}
\centerline{\footnotesize B. SCHWESINGER and H. WALLISER
}
\vspace*{0.015truein}
\centerline{\footnotesize\it FB Physik, University Siegen, Postfach 101240}
\baselineskip=12pt
\centerline{\footnotesize\it  D-57068 Siegen}
\vglue 12pt
\baselineskip=12pt
\baselineskip=6mm
\vglue 12pt
\abstracts{We derive an analytic expression for the Kroll-Ruderman
amplitude up to ${\cal O}(N_C^{-1})$
for general Skyrme-type models of the nucleon.
Due to the degeneracy of intermediate $N$- and $\Delta$-states we find
deviations from the standard low-energy theorem for the photoproduction
of neutral pions.}{}{}

\vspace{4.truein}

\centerline{\footnotesize SI-93-TP3S2}
\eject

\section{Introduction.}
At low energies the amplitude ${\cal F }$ for the production of
pions, $\bpi$, from
a photon $ a_\mu \sim \epsilon_\mu e^{i k \cdot x}$
incident on a nucleon\cite{CGLN}
arises from the contributions of  S-wave $\pi N$-channels.
The charge dependence of the reaction follows from three
independent amplitudes $E^{(-,0,+)}_{0^+}$:
\ba
{\cal F} =
\left\{-i (\bpi \times \btau)_3 E^{(-)}_{0^+}
+\bpi \cdot \btau E^{(0)}_{0^+}
+\bpi_3 E^{(+)}_{0^+}
\right\} i \bepsilon \cdot \bsigma ,
\ea
where the spin and isospin dependence of the nucleonic degrees of
freedom is expressed by Pauli-matrices $\sigma_k$ and $\tau_a$.
The amplitude  ${\cal F }$ is defined such that its  matrixelements
between the initial and final spin-isospin state
of the nucleon $ \mid i \rangle $, $ \mid f \rangle $, and the final isospin
state  $ \mid  \alpha \rangle $ of the pion, lead to the
differential reaction cross section
\ba
\frac{d \sigma_{c.m.}}{d \Omega_\pi} =
\frac{\mid q_\pi \mid }{ \mid k_\gamma \mid } \frac{1}{4}
\sum_{{\rm pol }} \mid \langle f , \alpha \mid {\cal F} \mid i \rangle \mid ^2.
\ea

Current algebra and PCAC fix the first terms of an expansion of the three
S-wave amplitudes
with respect to the coefficient of the chiral symmetry breaking,
i.e. the pion mass squared. This expansion, known as the
Kroll-Ruderman theorem\cite{KR,dB}, reads
\ba
E^{(-)}_{0^+}&\!\!\!=&\!\!\!\frac{\mid e \mid }{4 \pi} \frac{ g_A}{2 f_\pi}
C\left(\frac{m_\pi}{M}\right)
\left[ 1 + {\cal O}\left( (\frac{m_\pi}{M})^2 \right) \right]
\nonumber \\
E^{(0)}_{0^+}&\!\!\!=&\!\!\!\frac{\mid e \mid }{4 \pi} \frac{ g_A}{2 f_\pi}
C\left(\frac{m_\pi}{M}\right)
\left[ -\frac{1}{2} \frac{m_\pi}{M} +\frac{1}{4}(\mu_p+\mu_n)
(\frac{m_\pi}{M})^2 + {\cal O}\left( (\frac{m_\pi}{M})^3 \right) \right]
\\
E^{(+)}_{0^+}&\!\!\!=&\!\!\!\frac{\mid e \mid }{4 \pi} \frac{ g_A}{2 f_\pi}
C\left(\frac{m_\pi}{M}\right)
\left[ -\frac{1}{2} \frac{m_\pi}{M} +\frac{1}{4}(\mu_p-\mu_n)
(\frac{m_\pi}{M})^2 + {\cal O}\left( (\frac{m_\pi}{M})^3 \right) \right]
\, . \nonumber
\ea
For later convenience we have grouped factors of kinematical origin into
\ba
C\left(\frac{m_\pi}{M}\right)=
\frac{1+\frac{1}{2} \frac{m_\pi}{M}
+ {\cal O}\left( (\frac{m_\pi}{M})^2 \right) }
 { \left( 1+ \frac{m_\pi}{M} \right)^{\frac{3}{2} }  } \, .
\ea

Rightaway, from the first attempt\cite{ES} already, it has been
clearly visible
that the Skyrme model follows the Kroll-Ruderman-theorem closely, at
least numerically. La\-ter, it was understood\cite{SWHH},
that the zeroth order term in the pion mass entering the isovectorial
$E^{(-)}_{0^+}$-amplitude actually follows analytically. Further numerical
in\-vesti\-ga\-tions\cite{H,SD} have confirmed this and shown, that the slope
of
the $E^{(0)}_{0^+}$-amplitude with respect to the pion mass is
of the size required, although ref.\cite{H} disagrees on the sign.
In the Skyrme model the isoscalar
$E^{(0)}_{0^+}$-amplitude originates from the Wess-Zumino-anomaly.

The third, the isovectorial $E^{(+)}_{0^+}$-amplitude, finally,
remained zero due to the adiabatic approximation to meson-soliton
scattering adopted in\cite{ES,SWHH,SD}. In an $N_C$-counting scheme this
amplitude is down by one order relative to the other isovectorial
amplitude, because the nucleon mass $M$ is ${\cal O}(N_C)$ and the pion
mass $m_\pi$ is of order one. Thus this amplitude can only arise once
rotational effects of the soliton are taken into account in soliton-meson
scattering\cite{W}.

The purpose of the present work is to derive
analytic expressions for all three amplitudes complete up to the order
${\cal O}(N_C^{-1})$ relative to the leading terms in eq.(3).
This derivation turns out to be possible for the very general
class of chirally symmetric actions atmost quadratic in the time derivatives
of the meson fields. The result is different from the conclusions in
refs.\cite{H,SD,STU} which mutually disagree with each other.

\section{Low-energy $U$-matrix and current algebra}

In Skyrme-type models, the nucleon is based on
the field configurations of the hedgehog soliton
\be
U_H(\br)= e^{i \sbtau \cdot \hat r \chi(r) } \,
\ee
and it acquires its kinematical and spin degrees of freedom by
introduction of collective coordinates. For small velocities such
coordinates originate from a
Galilean transformation of the center of mass ${\bf X}(t)$
and from adiabatically slow rotations $A(t)$
\be
U = A\,  U_H(\br+{\bf X}) \, A^\dagger , \qquad
A^\dagger \dot A = -\frac{i}{2} \btau \cdot {\bf \Omega }
\rightarrow 0 \, .
\ee
of the soliton.

It was soon noticed\cite{KVW} that the rotational and translational
degrees of freedom in eq.(6) alone violate the
commutation relations of current algebra
\be
[Q_a,Q_b]=i\epsilon_{abc}Q_c\,,\qquad [Q_a,Q^5_b]=i\epsilon_{abc}Q^5_c
\,,\qquad
[Q^5_a,Q^5_b]=i\epsilon_{abc}Q_c
\ee
between the vector and axial charges
\be
Q_a=\int V^0_a(U) d^3r \, ,\qquad Q^5_a=\int A^0_a(U) d^3r \, .
\ee
As usual, the charges are defined as integrals over the time-components
of vector and axial vector currents
\be
V_a^\mu (U) = -i \trom \frac{\delta {\cal L}}{\delta \partial_\mu U}
[ \frac{\tau_a}{2} , U ] \, , \qquad
A_a^\mu (U) = -i \trom \frac{\delta {\cal L}}{\delta \partial_\mu U}
\{ \frac{\tau_a}{2} , U \} \, .
\ee
In seeming contradiction to this, it also became clear that
Skyrme-type models do reproduce
low-energy theorems, generally based on the current algebra relations,
such as the Tomozawa-Weinberg relation for S-wave
$\pi$N scattering\cite{UK,W} and the Adler-Weisberger sum rule\cite{UHS},
once ${\cal O}(N_C^{-1})$ rotational effects are properly taken into
account. Since the current algebra relations conventionally are the starting
point also for our present objective, the low-energy theorem of
photoproduction, we will reexamine the
commutation relations in eq.(7) taking ${\cal O}(N_C^{-1})$ effects into
consideration.

In order to describe $\pi$N scattering the ansatz,  eq.(6), must be
augmented by small amplitude fluctuations around the soliton. In the
low-energy region of interest here the configurations with fluctuations
may simply be written as
\be
U = A\, e^{\frac{i}{2f_\pi} \sbtau \cdot {\bf e}} \,
U_H(\br + {\bf X}+\frac{3 g_A}{2 f_\pi M} {\bf e} ) \,
e^{\frac{i}{2f_\pi} \sbtau \cdot {\bf e}} \, A^\dagger \, .
\ee
The collective coordinates ${\bf X}$, ${\bf e}$ and the Euler angles
contained in $A(t)$ are independent variables.
The fluctuation corresponding to the first order term of $U$ with respect
to the parameters ${\bf e} $ represents a linear combination of a
chiral rotation of the hedgehog and a translation.
In case of chiral symmetry, i.e. when the chiral symmetry breaking
mass term
\be
{\cal L}^{(CSB)}= \frac{f_\pi^2 m_\pi^2}{4} {\rm tr} (U +U^\dagger -2)
\ee
is absent, both modes are zero frequency solutions to the adiabatic
equations of motion for small amplitude fluctuations. The special
linear combination given here is determined by the fact, that
the S-wave scattering solution is orthogonal on the
localized, purely translational zero mode\cite{W}. The overlap integrals
between chiral rotations and translations involve the mass $M=-L[U_H]$
of the soliton and its axial coupling $g_A$,
\ba
\int A^j_a\left( A U_H A^\dagger \right) d^3r
= D_{ab}(A) \int A_b^j \left( U_H \right) d^3r =
-\frac{3}{2} D_{aj}(A)  \, g_A \, ,
\ea
independent of any specific choice of the total action $L$ or any
specific parametrization of the fluctuations.
Different parametrizations of the fluctuations lead to different
norm-kernels which
assure that the overlaps are always given in terms of mass and axial
coupling constant.

The $D$-Functions  $D_{ab}(A) = \frac{1}{2} \trom \tau_a A \tau_b
A^\dagger$ transform from soliton-fixed to physical isospin-axes such
that the physical pion fields $\bpi$ are related to the soliton-fixed
chiral rotation angles ${\bf e}$ via
\be
e_b=\pi_c D_{cb}(A) \, .
\ee

In the presence of the chiral symmetry breaking the changes of the
fluctuations are all ${\cal O}(m_\pi^2) $ except at low energies. There,
the time dependence of the fluctuation shifts to
\be
\bpi = {\bf a}_0 e^{-im_\pi t} + {\bf a}_0^\dagger e^{im_\pi t}
\ee
because the chiral symmetry breaking changes the asymptotic dispersion
relation of the fluctuations. Thus, even with broken chiral symmetry
the matrices in eq.(10) represent the exact low-energy
behaviour of the chiral fields up to ${\cal O}(m_\pi) $ apart from the
modifications to be made concerning the time dependence of the
fluctuations.

When one abandons the assumption of adiabaticity of pion-soliton scattering,
to lowest order the changes in the fluctuations of order ${\cal
O}(N_C^{-1})$ are driven by an inhomogeneous term linear
in the rotational velocities ${\bf \Omega}$ and linear in the adiabatic
fluctuation\cite{W}. Later, we
will transform the photoproduction amplitudes to a form where the
inhomogeneity can be inserted directly. Thus, up to order
${\cal O}(N_C^{-1})$  and ${\cal O}(\frac{m_\pi}{M})$ the general
structure of the chiral fields
in eq.(10) together with their time dependence in eq.(14) are sufficient
for the low-energy photoproduction amplitude.

As a consequence of the specific linear combination chosen in eq.(10) the
translation decouples from the other modes in the collective
lagrangian which up to order ${\cal O}(N_C^{-1})$ reads
\be
L=-M+\frac{1}{2} E \dot {\bf e}^2+\frac{1}{2} M \dot {\bf
X}^2+\frac{1}{2} \Theta {\bf \Omega}^2+ \left(E-\frac{\Theta}{2f^2_\pi}
\right){\bf \Omega} \cdot (\bf e \times \dot {\bf e}) \, .
\ee
The constant $E$ represents the infinite norm of the scattering state and
will be of no further significance. The rotation-vibration coupling of
order ${\cal O}(N_C^{-1})$, last term in eq.(15), involves the moment of
inertia $\Theta$ and
directly leads to the Tomozawa-Weinberg split of the S11 and S31
scattering lengths\cite{S}. The collective lagrangian, eq.(15), fixes the
conjugate momenta and angular momenta
\ba
\bf p \!\!\!&=&\!\!\!\frac{\partial L}{\partial \dot {\bf e}}=E \dot{\bf e}
+\left(E-\frac{\Theta}{2f^2_\pi} \right)\bf \Omega \times {\bf e} \nonumber \\
\bf R \!\!\!&=&\!\!\!\frac{\partial L}{\partial \bf \Omega}=\Theta {\bf \Omega}
+\left(E-\frac{\Theta}{2f^2_\pi} \right)\bf e \times \dot {\bf e} \\
\bf P \!\!\!&=&\!\!\!\frac{\partial L}{\partial \dot {\bf X}}=M \dot {\bf X}
\, .\nonumber
\ea
The angular velocities expressed by the conjugate momenta
\be
{\bf \Omega} = \frac{1}{\Theta} (\bf R - \bf e \times \bf p)
\ee
are of order ${\cal O}(N_C^{-1})$.
The vector and axial vector charges expressed in terms of the
collective momenta, eq.(16),
\ba
Q_a\!\!\!&=&\!\!\!D_{ab} \left[ \Theta {\bf \Omega} +
\left( E-\frac{\Theta}{2f^2_\pi} \right) {\bf e} \times
\dot {\bf e} \right]_b =D_{ab}R_b \nonumber \\
Q^5_a\!\!\!&=&\!\!\!D_{ab} \left[-f_\pi E \dot {\bf e} + \frac{3}{2} g_A
\dot {\bf X} -f_\pi \left(
E-\frac{\Theta}{f^2_\pi} \right) {\bf \Omega}
\times {\bf e} \right]_b  \\
   &=&\!\!\!D_{ab} \left[ -f_\pi {\bf p} +\frac{3}{2} \frac{g_A}{M}
{\bf P} +\frac{1}{2f_\pi} ({\bf R}-
{\bf e} \times {\bf p})\times {\bf e} \right]_b \nonumber \,
\ea
may now be used to verify the current algebra.
Cubic terms in the parameters $\bf e$ have been neglected, since
they will not enter
the relevant photoproduction amplitudes. With this the low-energy
charges in eq.(18) are correct next to leading order in $1/N_C$.
Postulating canonical quantization rules,
$[p_a,e_b]=-i\delta_{ab}\,$, $[P_a,X_b]=-i\delta_{ab}\,$,
$[R_a,R_b]=-i\epsilon_{abc} R_c\,$,
$[R_a,D_{bc}]=-i\epsilon_{ace} D_{be}\,$, these charges become
operators which should be hermitized properly. For these hermitean
operators it is then straight-forward to verify the commutation
relations in eq.(7).
Note that the non-adiabatic term in the axial charge, eq.(18),
is indispensible,
because, if neglected, $[Q^5_a,Q^5_b]=0$ follows immediately. Thus, the
ansatz (10) is in accordance with current algebra if ${\cal O}(N_C^{-1})$
contributions are taken properly into account. Since the pure
translations will not contribute to the photoproduction amplitudes
later we omit the collective coordinate ${\bf X}$ from now on.

\section{The photoproduction amplitude up to ${\cal O}(N_C^{0})$.}

The photoproduction amplitude ${\cal F }$ for the creation of one pion after
absorption of one photon to lowest order follows from the linear
photon vertex
\ba
L_\gamma = L_\gamma^V + L_\gamma^S =
-\mid e \mid \int a_\mu \left [ V^\mu_3(U) + \frac{1}{2} B^\mu(U)
\right ]_{lin} d^3r
\ea
where the chiral fields in the vector current $V_3^\mu(U)$ and the
winding number current $B^\mu(U)$  must be expanded up to linear order in
the fluctuations around the soliton\cite{ES}, i.e. up to linear order in
${\bf e}$. At pion threshold in the
c.m. system kinematical and phase space factors relating the
matrixelements of this interaction, eq.(19), to the amplitude
${\cal F}$, eq.(1), combine to $C\left(\frac{m_\pi}{M}\right) $, eq.(4),
if the photon field is normalized to
${\bf a} = -\frac{1}{4 \pi} \bepsilon e^{i k \cdot x}$. So we may
directly compare the interaction from eq.(19) with the Kroll-Ruderman
theorem omitting the factor $C$ in the latter, eq.(3).

In Coulomb gauge, $a_\mu=(0,-{\bf a})$, where the expansion of the
low-energy amplitudes in orders of $1/N_C$ is straight-forward,
we find
that the spatial part of the winding number current already requires one
time derivative on the chiral fields and thus will be of ${\cal
O}(m_\pi)$ or ${\cal O}(\Omega )={\cal O}(N_C^{-1})$. To lowest order
we therefore only have contributions from the vector current. For
the configurations under consideration, eq.(10), we use an identity
for the vector currents of a chirally rotated configuration\cite{SWHH}
which is an analogue of the current algebra relations.
Taking proper care of the shifted arguments of the hedgehog in eq.(10)
the identity reads here
\ba
V^\mu_3 \left( U \right) &\!\!\! = &\!\!\!
D_{3a}(A) \left[ {\bf V}^\mu \left( U_H(\br + \frac{3 g_A}{2f_\pi M}
{\bf e} ) \right) \right. \\
& & \left. \qquad \qquad -  \frac{1}{f_\pi}{\bf e} \times {\bf A}^\mu
\left( U_H(\br + \frac{3 g_A}{2f_\pi M} {\bf e}) \right) \right]_a
    + \left\{ {\rm terms~with~} \dot{\bf e}\, ,\dot{A} \right\}. \nonumber
\ea
Insertion of this expression into the photocoupling keeping the terms
linear in the fluctuations, i.e. linear in ${\bf e}$, with a photon
field normalized to ${\bf a} = -\frac{1}{4 \pi} \bepsilon e^{i k \cdot
x}$ immediatly
leads to the Kroll-Ruderman amplitude of order ${\cal O}(m_\pi^0)$, eq.(3),
\ba
L_\gamma^V \mid_{N_C^0} &\!\!\! = &\!\!\! -\mid e \mid D_{3a}
\int d^3r \, a_\mu
\left[ \frac{3 g_A}{2f_\pi M} {\bf e}\cdot \nabla
{\bf V}^\mu\left( U_H \right)
- \frac{1}{f_\pi}{\bf e} \times {\bf A}^\mu\left( U_H  \right)
\right]_a \nonumber \\
&\!\!\!=&\!\!\!-\frac{\mid e \mid }{8 \pi f_\pi} D_{3a}
( 3 g_A {\bf e} \times \bepsilon )_a =
\frac{\mid e \mid }{4 \pi} \frac{ g_A}{2 f_\pi}
(-i \bpi \times \btau )_3 i \bepsilon \cdot \bsigma \, .
\ea
Up to  order
${\cal O}(m_\pi^2)$ the translational part of the fluctuation doesn't
contribute  and  the integration of
the axial current of the hedgehog supplies the factor $g_A$, eq.(12).
We have used the substitution
$D_{ab} \rightarrow -\frac{1}{3}\tau_a \sigma_b $ for the
matrixelements of the $D$-function between nucleon states and replaced
the soliton-fixed fluctuation by its laboratory components, eq.(13).

\section{The isoscalar amplitude up to ${\cal O}(N_C^{-1})$.}

The adiabatic fluctuations inserted into the photocoupling from the
Wess-Zumino term leads to an isoscalar vertex which is of the order
${\cal O}(N_C^{-1})$ because the winding number
current
\be
B^\mu=\frac{1}{24 \pi^2} \epsilon^{\mu \nu \sigma \tau}
{\rm tr} (U^\dagger \partial_\nu U)( U^\dagger \partial_\sigma U)
(U^\dagger \partial_\tau U) \, .
\ee
is down by one order in $N_C$ relative to the isovector current.
The chiral rotation in the low-energy fluctuation only
contributes to the order ${\cal O }(m_\pi^2)$. The piece stemming from
the translation, on the other hand, will be proportional to $g_A$
because of the orthonormalization factors in eq.(10).
For winding number $B=1$ this piece immediately leads to the same
expression as in the Kroll-Ruderman theorem for $E^{(0)}_{0^+}$
, eq.(3), because the time dependence of a
pion in the final state, $\dot{ \bpi} = i m_\pi \bpi$, eq.(14),
must be inserted:
\ba
L_\gamma ^S \mid_{N_C^{-1}}    &\!\!\! = &\!\!\! -\frac{\mid e \mid }{2}
 \int a_j \frac{\epsilon^{0ijk}}{8 \pi^2} \frac{3 g_A}{2f_\pi M}
 \trom ( U_H^\dagger  \dot {\bf e} \cdot \nabla U_H
\, U_H^\dagger \partial_i U_H \, U_H^\dagger \partial_k U_H ) d^3r
\nonumber \\
&\!\!\! = &\!\!\! \frac{\mid e \mid }{8 \pi}
\frac{ 3 g_A}{2f_\pi M} (\dot{\bf e} \cdot{ \bepsilon }) B =
 \frac{\mid e \mid }{4 \pi}
\frac{  g_A}{2 f_\pi} (-\frac{m_\pi}{2M}  \btau \cdot  \bpi )
i \bsigma \cdot \bepsilon \, .
\ea

\section{The isovector amplitude in ${\cal O}(N_C^{-1})$.}

Up to this point the low-energy amplitude has been evaluated entirely in the
adiabatic approximation to meson-soliton scattering.
The soliton model has pieced different
factors entering the isoscalar amplitude together to an expression
identical to the one obtained by standard methods, but in a completely
different way.
The addition of non-adiabatic contributions now will
necessarily involve the rotational frequencies of the soliton and we
anticipate, that the little miracle that has happened in
case of the isoscalar amplitudes will continue to happen, i.e. the
resulting main corrections to the isovector amplitudes will turn out to be
entirely expressible in terms of $\mid e \mid $, $g_A$, $f_\pi$ and
$\frac{m_\pi}{M}$.

To demonstrate this we first transform the expression for the isovector
amplitude in eq.(19). The exact equations of motion for the
chiral fields are equivalent to the
vanishing of the divergence of the vector current
\ba
\partial_\mu {\bf V}^\mu (U) =\partial_0 {\bf V}_0 (U)
+\partial_i {\bf V}_i (U)=0  \, .
\ea
{}From this equation we write down the identity
\ba
\int a_i {\bf V}_i(U) d^3r
= \int a_j \partial_i\left( x_j{\bf V}_i(U) \right) d^3r
+\int a_j x_j \partial_0 {\bf V}_0(U) d^3r \, .
\ea
Let us first reconsider the case of adiabatic fluctuations.
By construction, the vector current linear in
the adiabatic fluctuations is divergenceless when all rotational
velocities are set to zero.
Therefore, the time derivative of the time component of the vector
current
contains two time derivatives both acting on the adiabatic fluctuation
such that at threshold the second term on the right hand side of eq.(25)
is already
${\cal O}(m_\pi^2)$. From the remaining term, upon partial integration,
the surface term at infinity just gives the contribution calculated in
the third section whereas the rest involves a derivative of the
photon field and may be discarded here: the spatial components
of the vector current linear in
the fluctuations only contain even multipoles in $\hat {\bf r}$ such that
the angular integration together with the derivative of the photon
field is at least quadratic in the photon momenta and thus ${\cal
O}(m_\pi^2)$.

If we now consider the non-adiabatic case then the first term on the
right hand side of eq.(25) does not add any new contributions: additions to the
surface term are zero because changes to the adiabatic chiral
rotation due to non-adiabatic terms vanish asymptotically and for the
same reasons as just outlined above the term with a derivative on the
photon field is at least ${\cal O}(m_\pi^2)$. Thus the non-adiabatic
contributions up to the order considered here only enter via the time
derivative of the time component of the vector current. Again, up to
the order considered, one of these time derivatives will furnish a
vibrational frequency $\omega = m_\pi$, the other one, necessarily, a
rotational frequency $\Omega$. So this term will be ${\cal
O}(N_C^{-1})$ with adiabatic fluctuations inserted and we may
safely drop higher order non-adiabatic corrections to the fluctuations.
With other words, we have
isolated the inhomogeneous term in the equations of motion which
drives the non-adiabatic
fluctuations and is proportional to the rotational velocity.

Expanded up to linear order in the adiabatic fluctuations ${\bf e}$
there are two
contributions to the time derivative of the time component of the
vector current, one from the global chiral rotation and one from the
orthonormalizing translation. The chiral rotation, using eq.(20), leads
to
\ba
V^0_3 \left( U \right) &=&
D_{3a}(A) \left[ \tilde {\bf V}^0_a \left( U_H \right)
- \frac{1}{f_\pi} \left( {\bf e} \times \tilde {\bf A}^0
\left( U_H \right) \right)_a \right. \\
&& \left. \qquad \qquad \qquad
-\frac{1}{f_\pi}\left( \dot {\bf e} + {\bf \Omega} \times {\bf e} \right)_c
\frac{\partial}{\partial \Omega_a} \tilde A^0_c \left( U_H \right)
\right ] , \nonumber
\ea
where we are using a somewhat sloppy notation for the body-fixed
currents of the rotating hedgehog which are linear in the rotational velocity:
\ba
D_{ab}(A) \tilde A_b^0\left( U_H \right) &=&  A_a^0\left( A U_H A^\dagger
\right) \\
D_{ab}(A)  \tilde V_b^0\left( U_H \right) &=&  V_a^0\left( A U_H A^\dagger
\right). \nonumber
\ea
Up to order ${\cal O}(\Omega ) \cdot {\cal O}(m_\pi)$ and linear in the
fluctuation we retain the terms
\ba
\partial_0 V^0_3 \left( U \right)\mid_{lin} &=&
-\frac{1}{f_\pi} D_{3a}(A) \left [
\left( {\bf \Omega } \times \frac{\partial}{\partial {\bf \Omega }}
\right)_a \dot {\bf e} \cdot \tilde {\bf A}^0\left( U_H \right) \right. \\
& & \left.\qquad + \left( \dot {\bf e} \times \tilde {\bf A}^0
\left( U_H \right) \right)_a
+\left( \ddot {\bf e} + {\bf \Omega} \times \dot {\bf e} \right)_c
\frac{\partial}{\partial \Omega_a} \tilde A^0_c \left( U_H \right)
\right] \nonumber
\ea
from the time derivative of the time component of the vector current.
Only the time component of the axial current of the
hedgehog enters into this expression, its general structure being
\ba
\tilde {\bf A}^0\left( U_H \right) = -\frac{3}{2} \theta (r) \, \cot \chi \,
\left[ {\bf \Omega } \times \hat r \right] \, .
\ea
The function $\theta$ is the angular averaged density for the
moments of inertia of the soliton. Due to this general structure of
the axial current and the relation
\ba
\ddot {\bf e } = - 2 {\bf \Omega } \times \dot {\bf e }
+ {\cal O}(\Omega ^2 ) + {\cal O}(m_\pi^2)
\ea
which follows from the equations of motion of the non-adiabatic
fluctuations or also from eq.(13,14) the time derivative in eq.(28), is
exactly zero
\ba
\partial_0 V^0_3 \left( U \right)\mid_{lin}&=&
-\frac{3 \theta}{2 f_\pi} \cot \chi \,D_{3a}(A) \left[
{\bf \Omega } \times \left( \hat {\bf r} \times \dot {\bf e} \right)
+\dot {\bf e} \times \left( {\bf \Omega } \times \hat {\bf r} \right)
\right. \\
& & \left. \qquad \qquad \qquad \qquad
-\hat {\bf r}  \times \left( {\bf \Omega } \times \dot {\bf e} \right)
\right]
= 0 \, . \nonumber
\ea

It remains to examine the contributions from the translation
which shifts the argument of the hedgehog configuration in eq.(10). Due to
this specific structure the linear terms of the isovector current with
respect to  ${\bf e }$ are easily
calculated:
\ba
L_\gamma ^V \mid_{N_C^{-1}}
&\!\!\!=&\!\!\!  -\frac{\mid e \mid }{4\pi} \int \br \cdot \bepsilon
\dot V^0_3(U)\mid_{lin,N_C^{-1}} d^3r
\\
&\!\!\! = &\!\!\! \frac{\mid e \mid }{4\pi}
\frac{3 g_A}{2 f_\pi M} D_{3a} \Theta \left(
4 \Omega_a \dot {\bf e}\cdot \bepsilon
-\epsilon_a {\bf \Omega } \cdot \dot {\bf e}
- {\bf \Omega } \cdot \bepsilon \dot e_a
\right) . \nonumber
\ea
The occurrence of the moments of inertia of the soliton, $\Theta$,
is directly related to the fact that the rotational energy is
proportional to
the integral of the time component of the vector current. In case of
time-dependent isospin rotations of a static soliton
this is true for any isospin symmetric action  atmost
quadratic in the time derivatives of the chiral fields.

Insertion of the physical pion-field, eq.(13,14),
and elimination of the angular velocities in favor of angular momenta
will now complete the derivation. However, at this point we are facing two
problems. The first one concerns the relation of the right angular momenta
to the angular velocities which up to order ${\cal O }(N_C^0)$ includes a
term bilinear in the
fluctuation and its conjugate momentum field\cite{W}.
Because of this extra term the action of the angular velocity on the
fluctuation ${\bf e}$ generates cubic terms in the mesonic fields.
In principle two of them could be contracted using the completeness
relation of the adiabatic fluctuations\cite{S}. From the
low-energy fluctuations in eq.(10) the piece given in eq.(17) may be
deduced. The unrestricted sum over intermediate scattering states on
the other hand leads to a multitude
of terms not expressible through
$\mid e \mid $, $g_A$, $f_\pi$ and $\frac{m_\pi}{M}$ alone. However,
all these terms
will necessarily be proportional to $D_{3a} \left( {\bf e} \times \bepsilon
\right)_a$, i.e. they provide ${\cal O }(m_\pi)$ corrections to the
$E^{(-)}_{0^+}$-amplitude. Here, we just retain
the term sufficient for the correct current commutators, eq.(17). Upon
hermitization its inclusion amounts to the replacement rule
\be
\Theta \Omega_a \dot e_b \rightarrow R_a \dot e_b + \frac{i}{2}
\epsilon_{abc} \dot e_c \,
\ee
for the angular velocity.

A second problem concerns ordering ambiguities related to the position
of the right angular momenta relative to two $D$-functions in the expressions
in eq.(32)
\ba
&&D_{3a} \Theta  \left(
4 \Omega_a \dot {\bf e} \cdot \bepsilon
-\epsilon_a {\bf \Omega } \cdot \dot {\bf e}
- {\bf \Omega } \cdot \bepsilon \dot e_a
\right)  =   \\
&&   \qquad             i m_\pi D_{3a} \left (
4 R_a \pi_c D_{cb} \epsilon_b  - \epsilon_a R_b \pi_c D_{cb}
 - \bepsilon \cdot {\bf R } \pi_c D_{ca} + \frac{5}{2} i(\bepsilon
\times {\bf e})_a
\right ) \, . \nonumber
\ea
Since ${\bf R }$ is a differential operator with respect to the Euler-angles
different orderings are distinguished by terms where ${\bf R }$
differentiates one of the two $D$-functions. The result of such a
differentiation is necessarily
of the form $D_{3a} \left( {\bf e} \times \bepsilon \right)_a$. Thus,
all uncertainties of the calculation presented here reside in the
${\cal O }(m_\pi)$ corrections to the $E^{(-)}_{0^+}$-amplitude.

The evaluation of the Euler-angle matrixelements will be given here
with respect to the ordering specified on the r.h.s. of eq.(34) which,
we think,
is actually the correct order: the differentiations apply to the
soliton-fixed fluctuation alone, just as in the case of the
inhomogeneous term in the equations of motion for the fluctuations.
Since the Euler-angle dependence of the laboratory fluctuation $\bpi$
must correspond to the final $\pi N$-channel, it is given by
$D^{(\frac{1}{2})}$-functions and the $\gamma N$ entrance channel
provides another  $D^{(\frac{1}{2})}$-function. Therefore we can use
the substitution $D_{ab} \rightarrow -\frac{1}{3}\tau_a \sigma_b $
for all the $D$-functions in eq.(34) where no couplings to
intermediate states of spins higher than $\frac{1}{2}$ are possible:
\ba
D_{3a} R_a \pi_c D_{cb} \epsilon_b &=& L_3 \pi_c D_{cb} \epsilon_b=
-\frac{1}{6} \left[ \bpi + i \bpi \times \btau \right]_3
\bsigma \cdot \bepsilon  \\
D_{3a} \epsilon_a R_b \pi_c D_{cb}&=&D_{3a} \epsilon_a L_c \pi_c =
-\frac{1}{6} \left[ \bpi + i \bpi \times \btau \right]_3
\bsigma \cdot \bepsilon \, . \nonumber
\ea
Note, that the left operators $L_a=D_{ab}R_b$ correspond to the isospin
carried by the Euler angles and the right operators to negative spin.
The third matrixelement in eq.(34) allows for
intermediate $\frac{3}{2}$-states in baryonic spin and isospin
\be
D_{3a}  \bepsilon \cdot {\bf R } \pi_c D_{ca} =
\left[D_{3a}, \bepsilon \cdot {\bf R }\right] \pi_c D_{ca}
+ \bepsilon \cdot {\bf R } \pi_3
=
\left[ -\frac{1}{2} \bpi  + \frac{i}{3} \bpi \times \btau\right]_3
\bsigma \cdot \bepsilon \, .
\ee

When we sum the four matrixelements together with the appropriate
coefficients
\be
L_\gamma ^V \mid_{N_C^{-1}}
=\frac{\mid e \mid }{4\pi}
\frac{g_A}{2 f_\pi}
\left[ 0 \cdot \pi_3
-0 \cdot i( \bpi \times \btau )_3
\right] i\bsigma \cdot \bepsilon \, ,
\ee
and we find a vanishing $E^{(+)}_{0^+}$-amplitude up to
${\cal O}(m_\pi)$ in disagreement with the low-energy theorem, eq.(3).
The vanishing correction to the $E^{(-)}_{0^+}$-amplitude, on the other hand,
is subject to several
uncertainties in the calculation, as has been discussed.

\section{Higher order corrections to the low-energy theorem}
In the standard formulation of the Kroll-Ruderman theorem, eq.(3),
it is possible to derive the
${\cal O}\left( (\frac{m_\pi}{M})^2 \right)$ corrections for the
$E^{(0)}_{0^+}$ and $E^{(+)}_{0^+}$ amplitudes. They are given in terms
of the anomalous magnetic moments of proton and neutron. Analogous
contributions may also be recovered explicitely from Skyrme-type
models. There are, however, other contributions to the same order
whose form cannot be given analytically and which
may not vanish, either.

We start with the isoscalar amplitude $E^{(0)}_{0^+}$. The contributions
of ${\cal O}\left( (\frac{m_\pi}{M})^2 \right)$ neglected till now
originate from two distinct cases:  the time derivative in the
spatial components of the winding number current will \\
(i) act on the
rotation matrices $A(t)$ and thus lead to a term of
${\cal O} ( N_C^{-2} )$ with adiabatic fluctuations inserted. Here each
factor $m_\pi$ is supplied by the photon momentum.\\
(ii) act on the non-adiabatic correction to the fluctuation which also
produces a term of ${\cal O} ( N_C^{-2} )$.\\
Case (i) allows an explicit derivation:
\ba
\delta L_\gamma ^S
&\!\!\!=&\!\!\!\frac{\mid e \mid }{8 \pi} \frac{3 g_A}{2f_\pi M}
 \int \epsilon_j e^{-i {\bf k} \cdot {\bf r}} {\bf e} \cdot \nabla
({\bf \Omega } \times \br)_j B^0(U_H) d^3r \nonumber \\
&\!\!\!=&\!\!\! \frac{\mid e \mid }{8 \pi}
\frac{ 3 g_A}{2f_\pi M} \frac{1}{3}\bepsilon \cdot ({\bf \Omega} \times {\bf
k } ) ({\bf e} \cdot{\bf k }) \langle r^2 \rangle \, .
\ea
The contribution from the pure chiral rotation vanishes locally leaving
the piece from the translation, only. The isoscalar mean square
radius is related to the isoscalar magnetic moment in hedgehog models
\ba
\mu^S = \mu_p + \mu_n = \frac{M}{3 \Theta} \langle r^2 \rangle,
\ea
thus the expression in eq.(38) immediatly leads to the standard result
\ba
\delta L_\gamma ^S
= \frac{\mid e \mid }{4 \pi}
\frac{  g_A}{2 f_\pi} \left[ \frac{1}{4}(\mu_p + \mu_n)
\frac{m_\pi^2 }{M^2}  \btau \cdot  \bpi \right]
i \bsigma \cdot \bepsilon
\ea
without any ordering ambiguities.

The translational part of the adiabatic low-energy fluctuation also
determines the isovectorial amplitudes of order ${\cal O}\left(
(\frac{m_\pi}{M})^2 \right)$. Again, each factor $m_\pi$ is supplied by
the photon momentum. Explicit, straight-forward calculation leads to
\ba
\delta L_\gamma ^V
=\frac{\mid e \mid }{8 \pi}
\frac{3  g_A}{2f_\pi M} \Theta D_{3a} ({\bf k } \times \bepsilon )_a
{\bf e } \cdot {\bf k }.
\ea
Now, the anomalous isovectorial magnetic moment appears because hedgehog
models always have
\be
\mu^V = \mu_p - \mu_n = \frac{2}{3} M \Theta .
\ee
The evaluation of the Euler-angle matrixelements of this expression
involves the steps
\ba
\pi_i k_b ({\bf k } \times \bepsilon )_a D_{3a} D_{ib} =
\frac{k^2}{2} \epsilon_{3ij} \pi_i D_{ja} \epsilon_a
= \frac{k^2}{6} i (\bpi \times \btau )_3 i \bsigma \cdot \bepsilon,
\ea
and we have used the orthogonality of the photon momentum on its
polarization. The resulting correction,
\ba
\delta L_\gamma ^V
=\frac{\mid e \mid }{4\pi}
\frac{g_A}{2 f_\pi}
\left[\frac{3}{8} \frac{m_\pi^2}{M^2}(\mu_p-\mu_n)
i( \bpi \times \btau )_3
\right] i\bsigma \cdot \bepsilon ,
\ea
resides in the wrong amplitude as compared to the standard expression, eq.(3).
However, the origin of the discrepancy is fairly easy to locate:
like in other cases\cite{K,BC}, the degeneracy of the
rotational states in the soliton model up to lowest order in $N_C^{-1}$
which allows nucleons and $\Delta$'s as intermediate states leads to
different expressions relative to the case where the $\Delta$'s are
excluded entirely from the calculation. We can actually implement the
second assumption by excluding
intermediate $\frac{3}{2}$-states in baryonic spin and isospin
but, unfortunately, this now will lead to ordering ambiguities with respect
to the $E^{(+)}_{0^+}$-amplitude. Intermediate
$\frac{3}{2}$-states are present in the matrixelement given in
eq.(43) and the immediate substitution $D_{ab} \rightarrow
-\frac{1}{3} \tau_a \sigma_b$ for all $D$-functions by matrixelements
of the $\frac{1}{2}$-representation eliminates contributions of the
higher spins. The ordering
\ba
\pi_i k_b ({\bf k } \times \bepsilon )_a D_{3a} D_{ib} \rightarrow
\frac{1}{9} \tau_3 \bsigma \cdot ({\bf k} \times \bepsilon )
\bpi \cdot \btau \bsigma \cdot {\bf k}
= \frac{k^2}{9} i \left( \bpi + i \bpi \times \btau \right)_3
i \bsigma \cdot \bepsilon
\ea
produces the standard correction to the $E^{(+)}_{0^+}$-amplitude
\ba
\delta L_\gamma ^V
=\frac{\mid e \mid }{4\pi}
\frac{g_A}{2 f_\pi}
\left[\frac{1}{4} \frac{m_\pi^2}{M^2}(\mu_p-\mu_n)
(\bpi + i \bpi \times \btau )_3
\right] i\bsigma \cdot \bepsilon
\ea
plus higher order corrections to the $E^{(-)}_{0^+}$-amplitude.
The latter
are not available in the standard form of the theorem, and thus we
have shown that we may bring the Skyrme model expressions into agreement with
the standard ones if we eliminate contributions of intermediate
$\Delta$-states.

Finishing this section we should emphasize once again, that there are
more corrections  to the
low-energy amplitudes, some of which would only be accessible numerically. The
first ones, ${\cal O}\left( (\frac{m_\pi}{M})^2 \right)$, arise in the
isoscalar amplitudes due to the contributions
from the case (ii), above. The others, already
${\cal O}\left( \frac{m_\pi}{M} \right)$,
originate, as discussed in section 5., from the relation between
angular velocities and right angular momenta. We only have
kept those terms that guarantee the correct commutation relations
between vector and axial charges up to ${\cal O} ( N_C^{-1} )$.
Latter uncertainties all reside in the
$E^{(-)}_{0^+}$-amplitude, the one which also suffers from ordering
ambiguities. Lastly, there are further corrections due to the implicit
dependence of the chiral angle on the chiral symmetry breaking. These
corrections may all be absorbed into the definition of $g_A$.

\section{Discussion and summary}
We have derived a low-energy theorem for the photoproduction of pions
on nucleons under no other assumption than "baryons are ridgidly rotated
solitons of a chirally invariant action atmost quadratic in the time
derivatives with a chiral symmetry breaking
of ${\cal O}(m_\pi^2)$". The unambiguous terms of this theorem in an
expansion in $\frac{m_\pi}{M}$ are summarized by
\ba
E^{(-)}_{0^+}&\!\!\!=&\!\!\!\frac{\mid e \mid }{4 \pi} \frac{ g_A}{2 f_\pi}
C\left(\frac{m_\pi}{M}\right)
\left[ 1 + {\cal O}\left( \frac{m_\pi}{M} \right) \right]
\nonumber\\
E^{(0)}_{0^+}&\!\!\! =&\!\!\! \frac{\mid e \mid }{4 \pi} \frac{ g_A}{2 f_\pi}
C\left(\frac{m_\pi}{M}\right)
\left[ -\frac{1}{2} \frac{m_\pi}{M}+\frac{1}{4}(\mu_p+\mu_n)
(\frac{m_\pi}{M})^2
+ {\cal O}\left( (\frac{m_\pi}{M})^2 \right) \right]
\\
E^{(+)}_{0^+}&\!\!\! =&\!\!\! \frac{\mid e \mid }{4 \pi} \frac{ g_A}{2 f_\pi}
C\left(\frac{m_\pi}{M}\right)
\left[ 0 \cdot \frac{m_\pi}{M}
+ {\cal O}\left( (\frac{m_\pi}{M})^3 \right) \right] ,
\nonumber
\ea
where the kinematical factors $C$ are given in eq.(4). The leading term
in the expansion of
the first two amplitudes, $E^{(-)}_{0^+}$ and $E^{(0)}_{0^+}$,
coincides with the standard low-energy
theorem\cite{dB,BKM}, the third amplitude, $E^{(+)}_{0^+}$, does not.

The origin of the discrepancy was suspected to reside in the degeneracy of the
rotational states in the soliton model up to lowest order in $N_C^{-1}$
which allows nucleons and $\Delta$'s as intermediate states.
Elimination of the contributions of the higher spins and the special ordering
\ba
\bepsilon \cdot {\bf R } D_{3a} \pi_c D_{ca} \rightarrow
\bepsilon \cdot {\bf R } \frac{1}{3} \tau_3  \bpi \cdot \btau
\ea
for the truncated matrixelement in eq.(36) would actually also reproduce the
standard prediction for the $E^{(+)}_{0^+}$-amplitude, but we could not
find any convincing justification for such a special ordering.
Corrections from a non-degenerate $\Delta $ to the photoproduction amplitudes
have already been considered a long time ago\cite{P} in the framework of
phenomenological lagrangians where the photocoupling of the $\Delta$
was introduced through an effective magnetic dipole operator. The couplings
taken from experiment have lead to corrections to the
low-energy theorem. However, the connection of this to the way soliton
models include higher rotational states remains obscure to us since
soliton models only have one local production vertex\cite{ES}.

Neither a vanishing nor an infinite nucleon-$\Delta$-split appear to be
realistic assumptions thus we do not see any compelling reason of why
one version of the low-energy theorem should be more realistic than the
other. Indeed, if we confront both with existing data we are not able
to find a contradiction to either version:
the reaction amplitudes in Walker's convention\cite{Walk} for specific
charge combinations
\ba
A^{n(\gamma , \pi^- ) p } &\!\!\!=&\!\!\! \sqrt{2}
(+E^{(0)}_{0^+}-E^{(-)}_{0^+} ) \nonumber \\
A^{p(\gamma , \pi^+ ) n } &\!\!\!=&\!\!\!
\sqrt{2} (-E^{(0)}_{0^+}-E^{(-)}_{0^+} )
\nonumber \\
A^{p(\gamma , \pi^0 ) p } &\!\!\!=&\!\!\!
(E^{(+)}_{0^+}+E^{(0)}_{0^+} )
\\
A^{n(\gamma , \pi^0 ) n } &\!\!\!=&\!\!\!
(E^{(+)}_{0^+}-E^{(0)}_{0^+} ) \nonumber
\ea
involve $E^{(+)}_{0^+}$ only in case neutral pions are produced.
Data for the production of $\pi^0$ on neutrons are not available. A
reanalysis of the data for the production of $\pi^0$ on protons vary
from\cite{Mainz} $A^{p(\gamma , \pi^0 ) p } = -(2.0 \pm .2) \cdot 10^{-3}
m_{\pi^+}^{-1}$ to\cite{Saclay} $-(1.5 \pm .3) \cdot 10^{-3}
m_{\pi^+}^{-1}$ as may be seen from table 1.,
where we confront the photoproduction  data with low-energy-theorems of
different origin.

Here, we concentrate on the production of neutral pions, where the soliton
model arrives at conclusions different from more standard approaches:\\
(i) Up to ${\cal O}(\frac{m_\pi}{M})$ the standard version, eq.(3),
predicts proton amplitudes which are too large
but the next order corrections, ${\cal O}((\frac{m_\pi}{M})^2)$, are
not small and lead to a number which is only slightly above the
Mainz data. The production amplitude on neutrons is predicted to be small.\\
(ii) The low-energy theorem has also been reconsidered recently in the
framework of chiral perturbation theory\cite{BKM} leading to differences in
${\cal O}((\frac{m_\pi}{M})^2)$ with respect to the standard expression.
Up to this order, in chiral perturbation theory the threshold
amplitude on protons has the wrong sign and only the full one loop
amplitude up to all orders in the pion mass leads to an amplitude
slightly below the Saclay data. The amplitude for the production on
neutrons is larger than the one on protons.\\
(iii) The soliton model only allowed a unique determination of the amplitudes
up to ${\cal O}(\frac{m_\pi}{M})$ giving amplitudes of equal magnitude
for the production of $\pi^0$ on protons or neutrons.
The amplitude for the production on protons lies between the Mainz and
the Saclay data.
Given the fact that the uncalculable next order still might lead
to substantial changes, no conclusions should be drawn, apart,
maybe, from the observation that the big differences between different
theorems seem to occur in the unmeasured amplitude for the production
of $\pi^0$ on neutrons.

The numbers in table 1. have been determined
by using the data everywhere for the masses $M$, $m_\pi$, the
electromagnetic charge and the $\pi N$-coupling constant $g_{\pi NN}$.
Since the Skyrme model also relates the axial charge to the $\pi N$-coupling
via the Goldberger-Treiman relation
\ba
g_{\pi NN}=M \frac{g_A}{f_\pi}+{\cal O}(m_\pi^2) ,
\ea
its replacement is correct up to the order indicated.
It is amusing to note that the apparent good numerical agreement of
Skyrme-type models concerning the photoproduction amplitudes can only
be obtained when data are inserted for the corresponding constants.
As is well known, no version of the Skyrme model can simultaneously fit
$\frac{g_A}{f_\pi}$ and $\frac{m_\pi}{M}$ unless the pion mass is
roughly doubled from its empirical value.

\begin {table} [h]
\tcaption{Kroll-Ruderman amplitudes  in units
$10^{-3} m_{\pi^+}^{-1}$ for the cases:
(i) the standard low-energy theorem, eq.(3),
(ii) chiral perturbation theory\cite{BKM}, CPT,
(q: up to quadratic order in
$m_\pi$, f: full one loop result),
(iii) soliton model according to eq.(47).
(iv) reanalysed experimental data, M: ref.\cite{Mainz}, S: ref.\cite{Saclay}.}

\begin{center}

\begin{tabular}  {| l| r r r | r |}
\hline
           & standard    &    CPT   & soliton  & experiment    \\
           &   LET~~~    &             &   model~ &        \\
\hline
& & & &   \\
$A^{n(\gamma , \pi^- ) p }$ & $-$31.8 & $-$31.5$^q$  & $-$31.8
                                              & $-$31.4$\pm$1.3$^M$ \\
                            &         & $-$31.1$^f$  &         & $-$32.2
                                                 $\pm$1.2$^S$ \\
                            &         &          &         &                 \\
$A^{p(\gamma , \pi^+ ) n }$ & $-$27.4 & $-$26.6$^q$  & $-$27.4 &
                                                         $-$27.9$\pm$0.5$^M$ \\
                            &         & $-$28.4$^f$  &         &
                                                        $-$28.8$\pm$0.7$^S$ \\
                            &         &          &         &                 \\
$A^{p(\gamma , \pi^0 ) p }$ &  $-$2.5 &     0.9$^q$  &  $-$1.6 &
                                                          $-$2.0$\pm$0.2$^M$ \\
                            &         &  $-$1.3$^f$  &         &
                                                          $-$1.5$\pm$0.3$^S$ \\
                            &         &          &         &          \\
$A^{n(\gamma , \pi^0 ) n }$ &   0.4   &     3.6$^q$  &     1.6 &       \\
                            &         &     3.6$^f$  &         &         \\
           &           &           &            &             \\
\hline
\end{tabular}
\end{center}

\end{table}

\section{Acknowledgement}
We thank N. Kaiser and U.-G- Meissner for drawing our attention to
ref.\cite{P}.
\vspace*{\fill}
\eject

\end{document}